\documentclass[aps,prl,onecolumn,superscriptaddress]{revtex4}
\usepackage{graphicx}
\usepackage{latexsym}
\usepackage{amsmath}
\usepackage{lmodern}
\usepackage{color,soul}
\usepackage{ulem}

\newcommand{\BFA}{BaFe$_2$As$_2$}
\newcommand{\BFAP}{BaFe$_2$(As$_{1-x}$P$_{x}$)$_2$}
\newcommand{\BFCA}{Ba(Fe$_{1-x}$Co$_x$)$_2$As$_2$}
\newcommand{\bog}{$\varepsilon_{\rm B_{1g}}$}
\newcommand{\btg}{$\varepsilon_{\rm B_{2g}}$}
\newcommand{\ag}{$\varepsilon_{\rm A_{1g}}$}
\newcommand{\aga}{$\varepsilon_{\rm A_{1g,1}}$}
\newcommand{\agc}{$\varepsilon_{\rm A_{1g,2}}$}
\newcommand{\ea}{$\varepsilon_{[100]}$}
\newcommand{\eab}{$\varepsilon_{[110]}$}

\begin{document}

\title{Uniaxial-Strain Tuning of the Intertwined Orders in BaFe$_2$(As$_{1-x}$P$_{x}$)$_2$}

\author{Zinan Zhao\textsuperscript{\S}}
\affiliation{Center for Advanced Quantum Studies and Department of Physics, Beijing Normal University, Beijing 100875, China}

\author{Ding Hu\textsuperscript{\S}}
\email{dinghuphys@hznu.edu.cn}
\affiliation{School of Physics, Hangzhou Normal University, Hangzhou 311121, China}
\affiliation{Center for Advanced Quantum Studies and Department of Physics, Beijing Normal University, Beijing 100875, China}

\author{X${\rm\ddot{u}}$e Fu}
\author{Kaijuan Zhou}
\affiliation{Center for Advanced Quantum Studies and Department of Physics, Beijing Normal University, Beijing 100875, China}

\author{Yanhong Gu}
\affiliation{Beijing National Laboratory for Condensed Matter Physics, Institute of Physics, Chinese Academy of Sciences, Beijing 100190, China}

\author{Guotai Tan}
\author{Xingye Lu}
\affiliation{Center for Advanced Quantum Studies and Department of Physics, Beijing Normal University, Beijing 100875, China}

\author{Pengcheng Dai}
\email{pdai@rice.edu}
\affiliation{Department of Physics and Astronomy, Rice University, Houston, Texas 77005-1827, USA}

\begin{abstract}
An experimental determination of electronic phase diagrams of high-transition temperature (high-$T_c$) 
superconductors forms the basis for a microscopic understanding of unconventional superconductivity.  For most high-$T_c$ superconductors, 
the electronic phase diagrams are established through partial chemical substitution, which also induces lattice disorder.
Here we show that symmetry-specific uniaxial strain can be used to study electronic phases in iron-based superconductors, composed of two-dimensional nearly square iron lattice planed separated by other elements.
By applying tunable uniaxial strain along different high symmetry directions and carrying
out transport measurements, we establish strain-tuning dependent electronic nematicity, antiferromagnetic (AF) order, and superconductivity of {\BFAP} 
superconductor.  We find that uniaxial strain along the nearest Fe-Fe direction can dramatically tune the AF order and superconductivity,
producing an electronic phase diagram clearly different from the chemical substitution-induced one. 
Our results thus establish strain tuning as a way to study 
the intertwined orders in correlated electron materials without using chemical substitution.
\end{abstract}

\maketitle
\section{Introduction}
One of the hallmarks of unconventional high-transition temperature (high-$T_{\rm c}$) superconductors is their intriguing phase diagrams resulting from the complicated interplay amongst the spin, charge, orbital, and lattice degrees of freedom \cite{Fradkin,Keimer,Scalapino,Johnston,stewart2011,dai,Qimiao2016,nsr2014}. In most unconventional superconductors, superconductivity appears after chemical doping to 
their non-superconducting antiferromagnetic (AF) ordered parent compounds. While chemical doping introduces electron/holes needed to suppress AF ordering temperature $T_N$ 
and induce superconductivity, it also brings about lattice disorder that affects the intrinsic electronic/magnetic properties of the system. 
 In iron-based high-$T_{\rm c}$ superconductors (FeSCs) with nearly square lattice of Fe plane, because of the close energy scales of AF, nematic order, 
and superconductivity \cite{stewart2011,dai,Qimiao2016}, symmetry-specific uniaxial stress/strain are used to tune the AF, nematic, and superconducting
(C) phase \cite{Chu2010,Fisher2012,kuo2012,kuo2016,ikeda2018,zhaoyu2019,BFCA1,hicks_prx,Fisher2021,ding2018,ding2020}.  
Specifically, uniaxial tensile and compressive strains along the Fe-Fe bond direction in the Fe plane, which breaks the crystalline 
four-fold rotational symmetry in the $B_{2g}$ symmetry channel and couples with nematic phase \cite{Chu2010,Fisher2012}, can dramatically 
suppress superconductivity with less than one per cent of strain in optimally electron-doped Ba(Fe$_{1–x}$Co$_x$)$_2$As$_2$ \cite{BFCA1}. 
These results are consistent with the fact that electron-doped Ba(Fe$_{1–x}$Co$_x$)$_2$As$_2$ 
has separate second order nematic and AF phase transitions with the former occurring at temperatures above $T_N$ and having a nematic quantum critical point (QCP) near optimal superconductivity \cite{Fisher2021}.  In {\BFAP} \cite{Jiang2009,Shishido}, although transport measurements indicate a nematic/AF quantum critical point (QCP) near optimal doping ($x=0.3$) with a diverging nematic susceptibility [Fig. 1(a)] 
\cite{kasa10,BFAP_science_2012,Shibauchi2014,fisher2014,Shishido2010,feng2012,kuo2016}, systematic neutron scattering experiments suggest 
that coupled nematic/AF phase transitions persist up to optimal superconductivity and the transition from nematic/AF to paramagnetic
phase is weakly first order without a QCP \cite{Allred2014,ding2015}. Since the effect of disorder induced by P-doping in {\BFAP} is unknown, it is important to use uniaxial strain as a probe to tune the phase diagram of {\BFAP} near optimal superconductivity 
and compare the outcome with the strain effect in optimally electron-doped Ba(Fe$_{1-x}$Co$_x$)$_2$As$_2$ \cite{BFCA1}.

\begin{figure}[htbp]
   \center{\includegraphics[width=8cm]{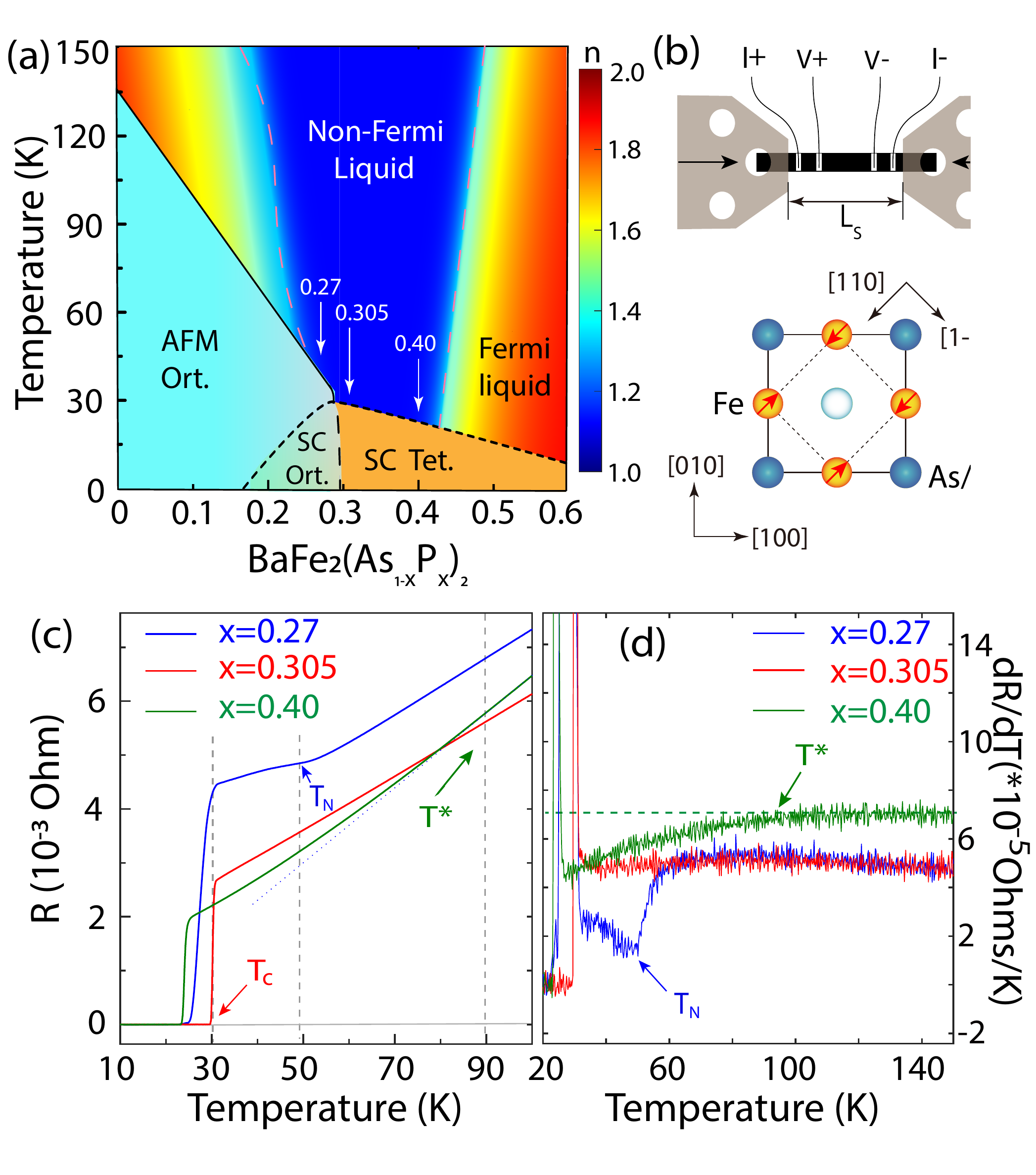}} 
   \caption{(a) Phase diagram of BaFe$_2$(As$_{1-x}$P$_x$)$_2$, where AF, SC, Ort. and Tet. represent antiferromagnetic, superconducting, orthorhombic, and tetragonal states respectively. The yellow (blue) areas mark the Fermi liquid (non-Fermi liquid) regime in the phase diagram \cite{kasa10,Shibauchi2014}. The doping levels used in this work are marked by the white arrows. (b) The upper panel illustrates the sample setup for the measurement of the strain-dependent resistivity. The lower shows a tetragonal unit cell in the FeAs/P plane, with the in-plane spin arrangement of the AF order depicted by the red arrows. (c) Temperature-dependent resistivity for $x=0.27, 0.305$, and $0.40$, and their (d) first-order derivatives $dR/dT$. The vertical dashed lines in (c) mark the $T_c$, $T_N$ and $T^*$, where  the $T^*$ marks the deviation of the resistance from T-linear dependence. }
   \label{fig1}
\end{figure}

In this paper, we report the uniaxial stress/strain tuning effects on superconductivity, nematic and AF orders of {\BFAP} through performing the stress/strain- and doping-dependent resistivity measurements on underdoped $x=0.27$ ($T_s= T_N \approx 49 $ K, $T_c\approx25$ K), optimally doped $x=0.305$ ($T_c\approx30$ K), and overdoped $x=0.40$ ($T_c\approx24$ K) samples [Figs.1(a-d)] \cite{Allred2014,ding2015}. 
We find that uniaxial strain along the Fe-Fe bond direction in the Fe plane is 
most effective in 
tuning the AF, nematic order, and superconductivity over a wide doping range. While uniaxial strain-tuned phase diagram is somewhat reminiscent that of P-doping dependent {\BFAP}, the range of superconductivity is greatly enhanced, establishing
 strain tuning as a way to study 
 the intertwined orders in {\BFAP} without the substitution induced lattice disorder.

\section{Results}
We start by describing resistivity measurements.  Uniaxial strains along the Fe-Fe bond direction and 45$^\circ$ degrees 
from it, corresponding to the $[1,1,0]$ and $[1,0,0]$ directions in the paramagnetic tetragonal phase, respectively \cite{Allred2014,ding2015}, were applied through a piezoelectric-stack driven commercial temperature-compensated strain cell (Razorbill CS130) with negligible thermal contraction on the bar-shaped single crystal \cite{hicks14,SRO1,SRO2}. 
The nominal longitudinal strain used in this work can be calculated via $\varepsilon_{xx}=\Delta L/L_s$, where $L_s$ denotes the sample length between the spacers fixing the two ends of the crystal, and $\Delta$L is the change of $L_s$ driven by the piezoelectric stacks  [Fig. 1(b)]  
and calibrated by the change of capacitance from the capacitive sensor in the cell \cite{SI}.
In this setup,  $\sim70 \%$ of the nominal strain can be transferred onto the crystal \cite{Fisher2021} and the stain distribution is homogenous between the two fixing spacers \cite{BFCA1}. 
To obtain temperature dependent resistivity curves without the influences from phase transitions, we applied a fixed voltage on the strain cell at 200 K before cooling down, then carried out resistance measurements while slowly heating up.
Since uniaxial strain along the $[1,1,0]$ direction can be decomposed to {\btg} and in-plane {\ag}
symmetry strain associated with the $D_{4h}$ point group of FeSC \cite{ikeda2018,BFCA1,hicks_prx,Fisher2021}, we can determine 
the roles of the strains {\ag}, {\bog}, {\btg} in tuning the intertwined orders.

 The application of $\varepsilon_{xx}$ will also cause a transverse strain $\varepsilon_{yy}=-\nu\varepsilon_{xx}$, and out-of-plane $\varepsilon_{zz}=-\nu'\varepsilon_{xx}$, resulting in the in-plane biaxial strain {\aga}$=\frac{1}{2}(\varepsilon_{xx}+\varepsilon_{yy})=\frac{1-\nu}{2}\varepsilon_{xx}$, in-plane uniaxial strain {\bog} (or \btg) $=\frac{1}{2}(\varepsilon_{xx}-\varepsilon_{yy})=\frac{1+\nu}{2}\varepsilon_{xx}$, and out-of-plane symmetric strain {\agc}=$-\nu'\varepsilon_{xx}$, where $\nu$ and $\nu'$ are the in-plane and out-of-plane Poisson's ratios, respectively. Figures 1(c) and 1(d) show the temperature-dependent resistivity and its first-order temperature derivatives ($dR/dT$) for the $x=0.27, 0.305$ and $0.40$ samples. The resistivity shows linear temperature dependence above $T_c$ for $x=0.305$, while deviates from the linear region (corresponding to non-Fermi liquid behavior) at $T^\ast\approx 56$ K for $x=0.27$, and $T^\ast\approx 90$ K for $x=0.40$ [Fig. 2(g-h)] \cite{Shibauchi2014, ding2018, ding2020}. In addition, the dip on the $dR/dT$ curve of the $x=0.27$ sample corresponds to the AF transition temperature and nematic phase transition with $T_N=T_s$ \cite{kasa10, Shibauchi2014}.

 \begin{figure*}[tphb!]
\includegraphics[width=17cm]{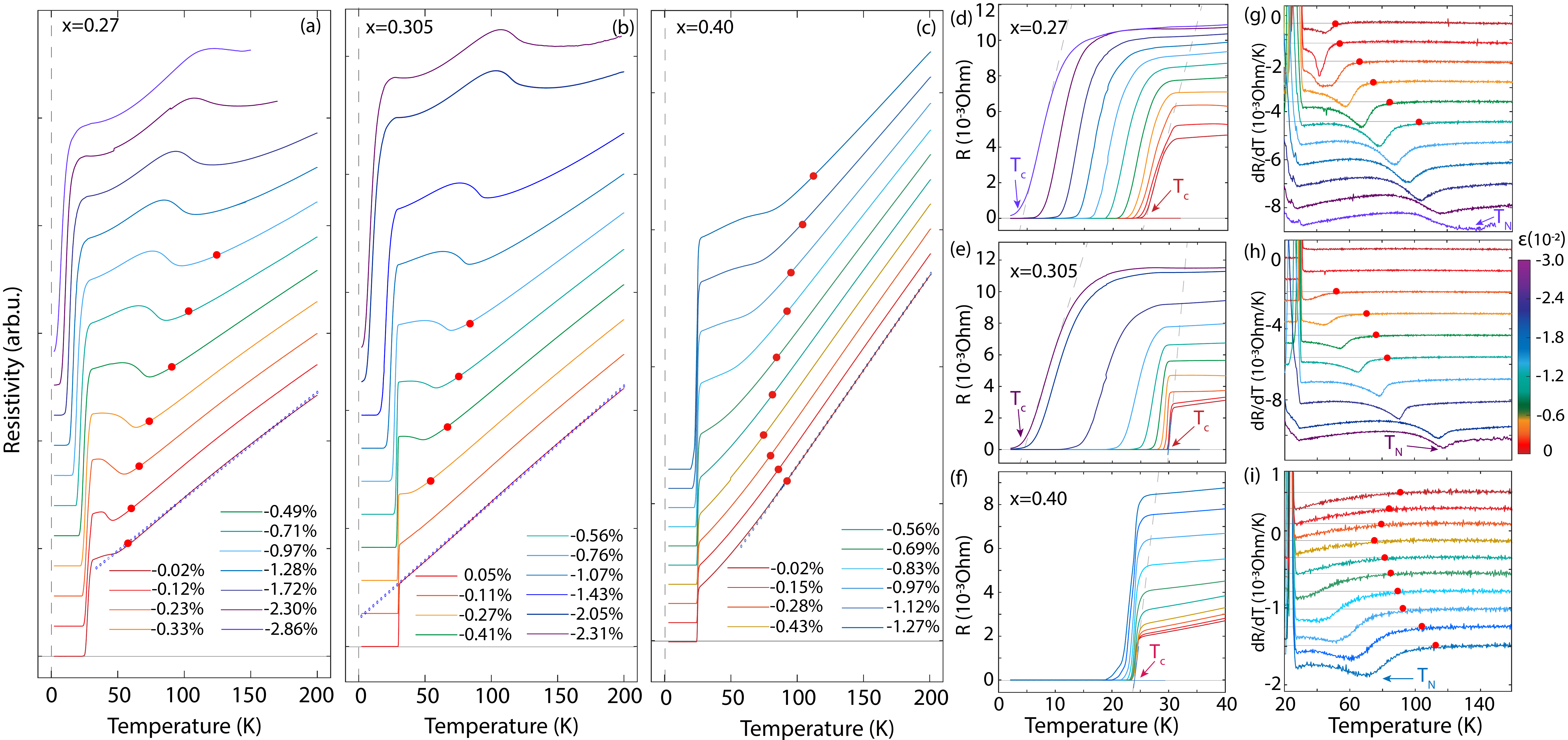}
\caption{(a-c) Temperature-dependent resistivity of the $x=0.27, 0.305$ and $0.40$ measured in a large temperature range under tunable uniaxial strains ($\varepsilon_{[110]}$)  along the $[1,1,0]$ direction. The data curves are shifted vertically for clarity. Dashed blue lines in (a-c) are linear fits for high temperature resistivity. Red dots mark the $T^*$. The lowest purplish lines are measured with zero voltage. (d-f) Temperature dependence of the resistivity below 40 K under $\varepsilon_{[110]}$ for the $x=0.27, 0.305$, and $0.40$. (g-i) Temperature-dependent of $dR/dT$ curves with different strains, where $T_N$ and $T^*$ are marked by arrows and red dots respectively.}
\label{fig2}
\end{figure*}

Temperature-dependent resistivity curves for the $x=0.27$, $0.305$, and $0.40$ samples under nominal uniaxial strains along the $[1,1,0]$ direction ($\varepsilon_{[110]}$) are shown in Figs. 2(a)-2(c), with the data below 40 K depicted in Figs. 2(d)-2(f), and $dR/dT$ in Figs. 2(g)-2(i), respectively. For underdoped $x=0.27$, we find that $\varepsilon_{[110]}$ can significantly enhance the $T_N$ [broad dip in Fig. 2(g)] and $T^*$ [red points in Figs. 2(a) and 2(g)]. By increasing $\varepsilon_{[110]}$ to  $-2.86 \%$, $T_N$ is enhanced to $\sim$120 K [Fig. 2(g)] close to the $T_N = 137 $ K for unstrained {\BFA} \cite{dai}. Superconductivity is simultaneously suppressed with increasing strain. By extrapolating the sharp drop of resistivity at the superconducting transition to zero resistance [Fig. 2(d)],
we find that $T_c$ is reduced to $4$ K at $\varepsilon_{[110]} \approx -2.86 \%$. For optimally doped $x = 0.305$ with $T_c = 30$ K, we also find strain-induced $T^*$  [red dots in Figs. 2(b) and 2(h)] and $T_N$ with $|\varepsilon_{[110]}| \geq$ 0.27 $\%$ [Fig. 2(h)]. With increasing {\eab}, $T_N$ and $T^*$ change to $\sim$120 K and $T_c$ is suppressed to around 4 K [Fig. 2(e)], before the crystal broken under $|\varepsilon_{[110]}| > $2.31 $\%$ [Fig. 2(b)]. In overdoped regime, 
the $x = 0.40$ sample has similar $T_c$ to that of $x=0.27$ but no AF and nematic order [Figs. 2(c) and 2(f)].
The linear temperature dependence of resistivity is suppressed to a lower temperature $T^* \approx 75$ K at $\varepsilon_{[110]}$ = -0.43 $\%$ strain and enhanced to $\sim 110$ K at larger strain. At $|\varepsilon_{[110]}| \geq 0.83 \%$, a much broader dip in $dR/dT$ could be attributed to the appearance of AF order, similar to the case for the $x = 0.305$ sample [Fig. 2(i)]. The induced $T_N$ of $x=0.40$ can be pushed to  $\sim70$ K at $\varepsilon_{[110]} = -1.27 \%$ [Fig. 2(i)]. However, there is relatively small strain-induce variation of $T_c$ in the overdoped sample [Fig. 2(f)].

\begin{figure}
\includegraphics[width=8.5cm]{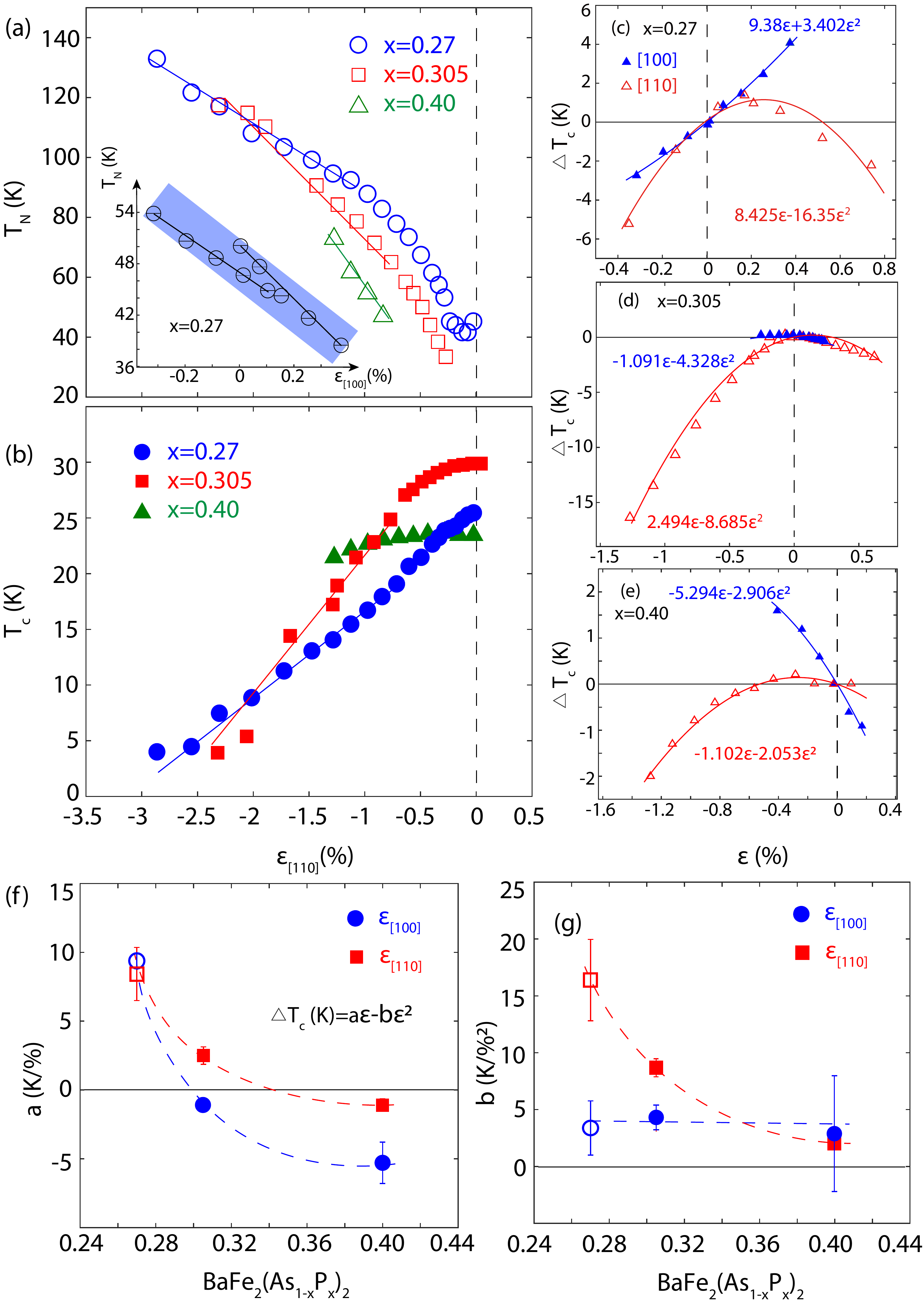}
\caption{(a,b) $\varepsilon_{[110]}$ dependence of $T_N$ and $T_c$ for the $x=0.27, 0.305$, and $0.40$. The inset of (a) is {\ea}-dependence of $T_N$ in {\ea}$<0.37\%$ for the $x=0.27$.  (c-e) {\ea}- and {\eab}-dependence of $\Delta T_c$ for the $x=0.27$, $0.305$ and $0.40$ samples. The blue and red lines are the fitting results of $\Delta T_c$({\ea}), and $\Delta T_c$({\eab}) with Eq. (1), respectively. The vertical dashed lines in (a-e) mark the zero strain point. (f,g) Doping dependence of the fitting parameters (f) $a$ and (g) $b$ for $\Delta T_c$({\ea}) and $\Delta T_c$({\eab}). Points of $x=0.27$ compound are marked by hollow points. The blue and red dashed curves are guides to the eyes.  }
\label{fig3}
\end{figure}

Figures 3(a) and 3(b) summarize $\varepsilon_{[110]}$-dependent $T_N$ and $T_c$ for $x=0.27, 0.305$ and $0.40$, obtained from the data shown in Fig. 2. For the $x=0.27$ and $0.305$ samples, the overall $\varepsilon_{[110]}$ dependence of $T_c$, $T_N$ and $T^*$ are akin to the doping dependence of $T_c$ and $T_N$ in the phase diagram of {\BFAP} at ambient pressure. However, as $\varepsilon_{[110]}$ will induce $\varepsilon_{[1\bar10]}=-\nu\varepsilon_{[110]}$ and $\varepsilon_c=-\nu'\varepsilon_{[110]}$, it can be decomposed to antisymmetric strain {\btg}$=\frac{1+\nu}{2}\varepsilon_{[110]}$, and symmetric strains {\aga}$=\frac{1-\nu}{2}\varepsilon_{[110]}$ and {\agc}$=-\nu'\varepsilon_{[110]}$. In order to explore the mechanism of the uniaxial strain tuning effects, it is necessary to understand the roles of {\bog}, {\btg}, and {\ag} in driving the intertwined orders. In principle, this decomposition is valid in the $D_{4h}$ point group. Therefore, we applied the strains at 200 K, which is much higher than the tetragonal-to-orthorhombic structural transition. 
For the $ x=0.27$ sample with orthorhombic lattice distortion around 0.035$\%$ at $T_c$ under ambient condition \cite{ding2015}, the applied strains are significantly larger than needed for detwinning the crystal. Thus, we carried out the symmetry decomposition analysis on the $x=0.27$ sample and compared the results with the $0.305$ and $0.40$ samples.

Besides {\eab}-dependent $T_c$ and $T_N$, we have also measured resistivity for the $x=0.27, 0.305$ and $0.40$ under uniaxial strains along the tetragonal $[1,0,0]$ direction ($\varepsilon_{[100]}$), which can be decomposed to {\bog} and {\ag}. In low-strain range,  the strain dependence of $\Delta T_c$ and $\Delta T_N$ can be mathematically described by
\begin{eqnarray}
\Delta T_c(\varepsilon)=a\varepsilon-b\varepsilon^2\\
\Delta T_N(\varepsilon)=\alpha\varepsilon+\beta\varepsilon^2
\end{eqnarray}
for which {\ag} is linearly coupled to $\Delta T_c$ ($a\varepsilon=a_1\frac{2}{1-\nu}${\aga}$-a_2\frac{1}{\nu'}${\agc}) and $\Delta T_N$, while {\bog}/{\btg} contribute only to the quadratic term ($\varepsilon=\varepsilon_{\rm B_{1g}}$ for {\ea} and $\varepsilon_{\rm B_{2g}}$ for {\eab}) \cite{ikeda2018}, where $a$, $b$, $\alpha$ and $\beta$ are doping-dependent dimensionless coupling coefficients \cite{ikeda2018}. 

Figures 3(c)-3(e) show the uniaxial-strain-induced change of $T_c$ ($\Delta T_c$) under $\varepsilon_{[100]}$ and $\varepsilon_{[110]}$ \cite{SI}. 
$\Delta T_c(\varepsilon_{[100]}$) for $x=0.27, 0.305$ and $0.40$ follows a linear strain dependence (contribution from {\ag}), while $\Delta T_c(\varepsilon_{[110]})$ contains substantial contributions from quadratic term ({\btg}). To make quantitative analysis, we show in Figs. 3(f) and 3(g) the fitting parameters for $\Delta T_c(\varepsilon_{[100]})$ and $\Delta T_c(\varepsilon_{[110]})$ using Eq. (1). The coupling to {\ag} ($a$) in both cases decrease with the increasing doping and changes its sign across the optimal doping, suggesting its 
possible connection to the QCP in {\BFAP}. 

In previous resistivity and neutron diffraction measurements of {\BFAP} under $c$-axis pressure, $T_c$ was enhanced in the underdoped $x=0.28\ (T_c\approx28$ K) sample by $\Delta T_c/\Delta p_c\approx3$K/GPa ($\Delta T_c$ roughly depends on $p_c$ linearly below $\sim0.3$ GPa) \cite{ding2020},  but decreased in the optimally doped $x=0.30$ ($T_c\approx30$ K) by $dT_c/dp_c\approx-3$ K/GPa \cite{ding2018}. Since $\varepsilon_{[100]}$ induces {\aga} and {\agc}=-$\frac{2\nu'}{1-\nu}${\aga}, while a $c$-axis pressure generates compressive {\agc} and tensile {\aga}=$-\nu''${\agc} ($\nu''\approx0.47$ for $x=0.3$ determined in ref. \cite{ding2018}, $\nu''\approx\frac{2\nu'}{1-\nu}$) \cite{SI}, a $c$-axis pressure has similar effects to tensile $\varepsilon_{[100]}$, whose effects on $x=0.28$ and $0.3$ are therefore in line with the doping dependence of the coupling between {\ag} and $\Delta T_c$. This further constrains the sign reversal of $a$ to the optimal doping regime in {\BFAP}.

In contrast to the ${\rm A_{1g}}$ channel, $\varepsilon_{[100]}$ and $\varepsilon_{[110]}$ behave differently in their in-plane antisymmetric strains {\bog} and {\btg}. The coefficient $b$ for {\bog} is essentially doping independent and negligible in underdoped regime [Fig. 3(g)]. However, the efficiency of {\btg} in tuning $T_c$ decreases with the increasing doping and becomes comparable with $b_1$ in the overdoped $x=0.40$ sample [Fig. 3(g)].
Thus, the $\Delta T_c(\varepsilon_{[110]}$) (red solid triangles) for the underdoped regime ($x=0.27$) is driven by the collaborative {\ag} and {\btg}, but dominated by the {\btg} at the optimal doping regime. In overdoped regime, $T_c$ is very insensitive to {\ag}, {\bog} and {\btg}.

In resistivity measurements along the $[1,0,0]$ direction, $|\varepsilon_{[100]}|$ ($<0.37\%$) is found to tune the $T_N$ in the $x=0.27$ linearly by $dT_c$/$d\varepsilon_{[100]}$$\approx-21\pm5$ K/\% [inset of Fig. 3(a)], but fails to induce any distinguishable feature in resistivity corresponding to $T_N$ for $x=0.305$ and $0.40$ \cite{SI}. 
This suggests that {\ag} plays an important role while {\bog} is ineffective in tuning $T_N$ (and $T_s$), similar to previous results on {\BFCA} and FeSe \cite{ikeda2018,Fisher2021,hicks_prx}. 
Although $\varepsilon_{[110]}$ generates {\ag} with similar magnitude as $\varepsilon_{[100]}$, it induces AF order in both the optimally doped $x=0.305$ and overdoped $x=0.40$ samples and drive $T_N$ for the $x=0.27$ and $x=0.305$ samples to $\sim120$ K at $\varepsilon_{[110]}$ above $-2.3\%$. 
Moreover, the $\Delta T_N(\varepsilon_{[110]})$ for $x=0.27$ in the low-strain range $|\varepsilon|<0.37\%$ seems dominated by a quadratic term, which may arise from {\btg}  [Fig. 3(a)] \cite{SI} . Therefore, {\btg} plays a dominant role in tuning $T_N$ in both the underdoped and optimally doped regimes. 
For comparison, the efficiency ($\beta$) of {\bog} in tuning $T_N$ changes less from $x=0.27$ to $0.40$. 
Previous phenomenological Ginzburg-Landau model suggests that the increase in $T_N$ is proportional to the applied strain, resulting in a V shape strain dependence of $T_N$ \cite{kuo2012} .
Unfortunately, we cannot distinguish V shape from quadratically strain dependence based on our data.


\begin{figure}
\includegraphics[width=15cm]{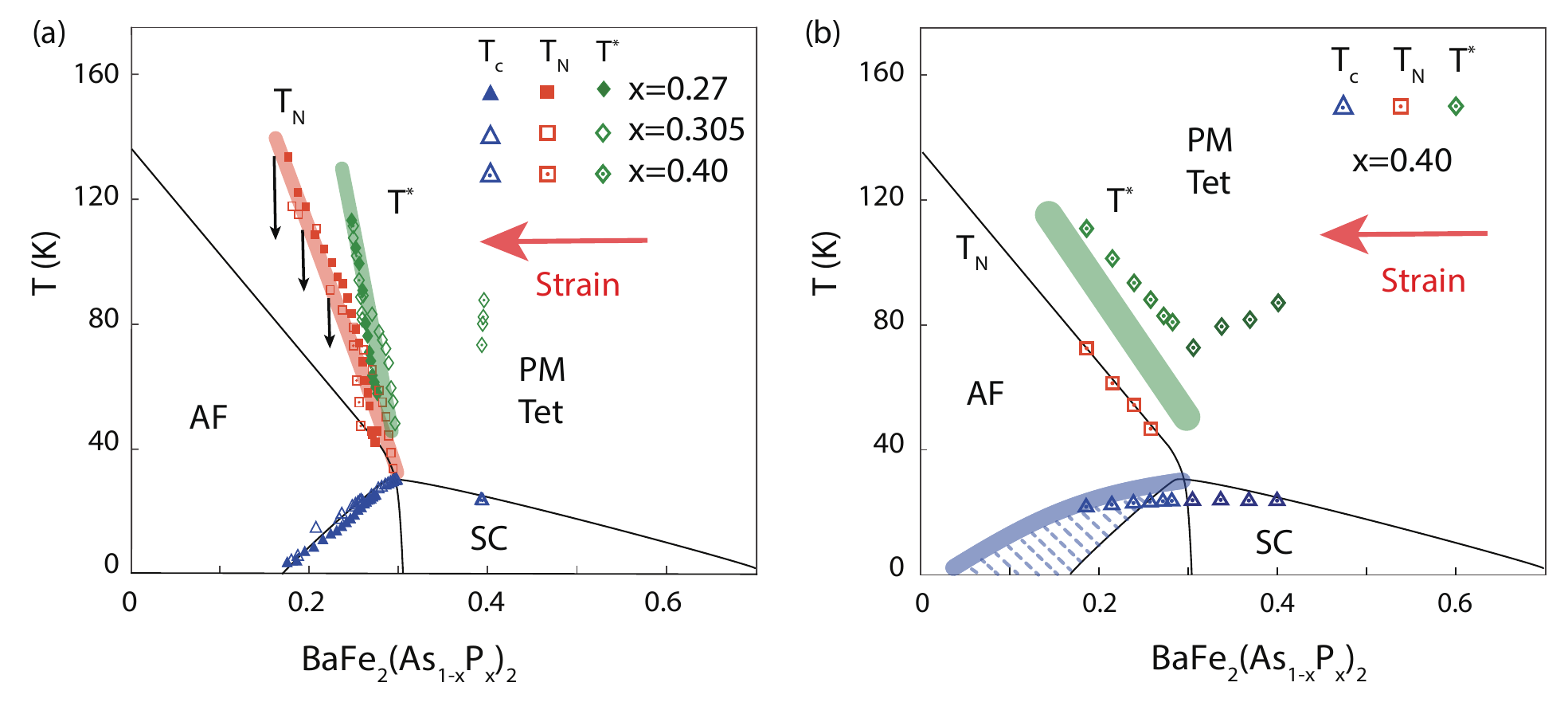}
\caption{(a) Phase diagram of {\BFAP} overlaid with the {\eab}-driven $T_N$, and $T^*$
 normalized by $T_{c}(\varepsilon_{\rm max})= T_c(x')$ and $T_{c}(\varepsilon=0)= T_c(x)$. The red arrows indicate the modulated direction of reproduced phase diagram by increasing $\varepsilon$. Black arrows show the amendatory direction of $T_N$ and $T^*$ due to the underestimate strains from $T_c$ normalization.  Thick pink and green lines are guides to the eyes. (b) Phase diagram normalized by $T_N$, and the points without  $T_N$ for x = 0.40 were normalized by $T^*_{min}=T^*(x = 0.305)$ and zero strain point. The blue shaded area is the extended superconducting regime in strain tuning phase diagram.
 Green and blue thick lines mark the evolution of $T^*$ and $T_c$ of $x=0.305$ and $0.40$ compounds.
 The black solid curves mark the $x$-dependent $T_N$ and $T_c$ in (a, b).}
\label{fig4}
\end{figure}

Having determined the impact of {\ag}, {\bog} and {\btg} in {\ea} and {\eab} in tuning $T_c$ and $T_N$, we compare the strain-induced phase diagrams to the doping-dependent phase diagram of {\BFAP} at ambient condition. 
To make a direct comparison and eliminate the influences from expansivity mismatch between crystal and Ti used in the cell, hysteretic behavior of piezo-stack, different transferred strain ratios in different measurements, $T_c$ or $T_N$ could be used to normalize the phase diagrams. 
By connecting {\eab} and $x$ through $T_c(x')=T_c(\varepsilon_{\rm max})$ and $T_c(x)=T_c(\varepsilon=0)$ that $T_N$ and $T^*$ were calibrated linearly \cite{Shibauchi2014,kasa10,fisher2014}, we overlaied the {\eab}-dependent transition temperatures onto the doping phase diagram in Fig. 4 (a).
Both for $x = 0.40$ and $0.305$ compounds, the initial induced $T_N$ points located right on the $T_N - x$ line measured at ambient condition. With the increasing of {\eab}, enhanced $T_N$ and $T^*$ of three compounds locates significantly above the $T_N - x$ line and evolve consistently, producing a universal electronic phase diagram induced by strain, wherever originating from underdoped, optimally doped or overdoped sides. 
The prominent difference between the {\eab} and $P$ substitution-induced phase diagram is their $T_N$'s under higher strain level.  It is worth noting that we have carried out temperature dependent resistance measurement over 2 K to 200 K to cover the large evolution ranges of $T_c$, $T_N$ and $T^*$. Partial distances would be attributed to the larger applied strains at $T_N$ due to the character of piezoelectric stacks with the adjusted directions marked by the arrows in Fig 4 (a).  
For the $x=0.40$ sample, compressive {\eab} drives the system across the QCP from the overdoped regime to the underdoped regime with induced $T_N$.
Due to the relatively small changes of $T_c$, it is convenient to reveal this process by $T_N$ normalized phase diagram as shown in Figure 4(b). The larger superconducting regime in this phase diagram reveals the possible advantage of strain modulation without effect from disorder.

\section{Discussion}

In summary, we have used uniaxial stress/strain as a way tune the phase diagrams of {\BFAP}.  We find that uniaxial strain along the $[1,1,0]$ direction ($\varepsilon_{[110]}$) is most effective in 
tuning the AF, nematic order, and superconductivity over a wide doping range, originating from a strong magnetoelastic coupling in the ${\rm B_{2g}}$ channel. 
The intertwined electronic phases of underdoped, optimally doped and overdoped compounds response to $\varepsilon_{[110]}$ universally. 

In electron-doped {\BFCA} with $x \geq 0.071$ (optimally doped level), superconductivity has also been suppressed without the revived magnetic order signal under $\varepsilon_{[110]}$, suggesting the appearance of enhanced spin fluctuations and suppressed nematic fluctuations. It associates the superconducting pairing to the nematic fluctuations instead of spin fluctuations  \cite{BFCA1}.
In contrast, the magnetic order 
revives in optimally doped  {\BFAP} ($x=0.305$) and has been enhanced to 120 K at the sacrifice of superconductivity. 
It refers to that not only nematic fluctuations, but also spin fluctuations have been suppressed by $\varepsilon_{[110]}$, suggesting their intimate positive correlation with superconductivity. It has also been confirmed in overdoped $x=0.40$ sample, where the $T_c$ start to be suppressed obviously when $T_N$ appears at $\varepsilon_{[110]}\geq 0.83 \%$ as shown in Figure 2. (f, i).
There is no evidence of separation between $T_N$ and $T_s$ even under $\varepsilon_{[110]} \textgreater 2 \%$, implying the strong influence from magnetoelastic coupling in {\BFAP}. 

In addition, remarkable T-linear resistivity in {\BFAP} has been believed to be derived from the existence of quantum fluctuations that dominating the physical properties above $T^*$. We found that $T^*$ can also been modulated by $\varepsilon_{[110]}$ without the existence of magnetic order in $x=0.40$ compound, forming a V shaped non Fermi-liquid region as shown in Figure 4(b).
Our results thus establish strain tuning as a way to study
the intertwined orders and related quantum fluctuations in {\BFAP}, but in an opposite direction compared with $P$ substitution and without the substitution induced lattice disorder.

The research at Beijing Normal University is supported by the National Key Projects for Research and Development of China (Grant No. 2021YFA1400400) and National Natural Science Foundation of China (Grant No. 11922402 and 11734002) (X.L.). The research at HangZhou Normal University is supported by the Open Project of Guangdong Provincial Key Laboratory of Magnetoelectric Physics and Devices (Grant No. 2022B1212010008), Startup Project of HangZhou Normal University (Grant No. 2020QDL026) and Natural Science Foundation of Zhejiang Province (Grant No. LY22A040009).
A part of the material synthesis/characterization work at Rice University is supported by the Robert A. Welch Foundation Grant No. C-1839  and the U.S. Department of Energy, BES under Grant No. DE-SC0012311 (P.D.).

\end{document}